# Global structure of thermal tides in the upper cloud layer of Venus revealed by LIR onboard Akatsuki

T. Kouyama[1]*, M. Taguchi[2], T. Fukuhara[2], T. Imamura[3], T. Horinouchi[4], T. M. Sato[5], S. Murakami[6], G. L. Hashimoto[7], Y. J. Lee[8], M. Futaguchi[9] T. Yamada[2], M. Akiba[2], T. Satoh[5], and M. Nakamura[5]

[1]Artificial Research Center, National Institute of Advanced Industrial Science and Technology, Tokyo, 165-0034, Japan

[2]College of Science, Rikkyo University, Tokyo, 171-8501, Japan

[3]Graduate School of Frontier Science, The University of Tokyo, Tokyo, 227-8561, Japan

[4]Graduate School of Environment Science, Hokkaido University, Sapporo, 060-0810, Japan

[5]Space Information Center, Hokkaido Information University, Ebetsu, 069-8585, Japan

[6]Institute of Space and Astronautical Science, Japan Aerospace Exploration Agency, Sagamihara, 252-5210, Japan

[7]Faculty of Science, Okayama University, Okayama, 700-8530, Japan

[8]Technische Universität Berlin, Berlin, 10623, Germany

[9]Omori medical center, Toho University, Tokyo, 143-8541, Japan

*Corresponding author: Toru Kouyama (t.kouyama@aist.go.jp)

**Key Points:**

- Akatsuki/LIR revealed the global structures of thermal tides across the equator in the upper cloud layer of Venus for the first time.

- Using the emission angle dependence of LIR's sensing altitude, upward propagation of the semidiurnal tide was confirmed.

- Wave types consisting of the thermal tides were identified.

**Plain Language Summary**

On Venus, the atmosphere circulates 60 times faster than the solid body of Venus; this phenomenon is called "superrotation", and it is one of the mysteries of the Venusian atmosphere. To maintain the fast circulation, thermal tides, which are global-scale atmospheric waves excited by solar heating, have been considered a very important candidate because they have ability of accelerating the atmosphere through propagating. A mid-infrared camera onboard the Japanese Venus orbiter, Akatsuki, can capture temperature perturbations due to the thermal tides in the upper cloud level (60-70 km altitude), and it revealed their global and vertical structures with a long-term observation (more than three Venusian years) for the first time. Interestingly, we found that the location of the maximum brightness temperature around the cloud top was different from noon where solar energy input is at a maximum. In addition, the location was shifted toward morning side as the sensing altitude increased. This finding is an evidence of the vertical travelling of the thermal tides, indicating the wave's atmospheric acceleration.

**Abstract**

Longwave Infrared Camera (LIR) onboard Akatsuki first revealed the global structure of the thermal tides in the upper cloud layer of Venus. The data were acquired over three Venusian years, and the analysis was done over the areas from the equator to the mid-latitudes in both hemispheres and over the whole local time. Thermal tides at two vertical levels were analyzed by comparing data at two different emission angles. Dynamical wave modes consisting of tides were identified; the diurnal tide consisted mainly of Rossby-wave and gravity-wave modes, while the semidiurnal tide predominantly consisted of a gravity-wave mode. The revealed vertical structures were roughly consistent with the above wave modes, but some discrepancy remained if the waves were supposed to be monochromatic. In turn, the heating profile that excites the tidal waves can be constrained to match this discrepancy, which would greatly advance the understanding of the Venusian atmosphere.

# 1 Introduction

Thermal tides excited in the cloud layer has been considered one of the main acceleration sources that maintain the atmospheric superrotation of Venus [e.g., Fels and Lindzen, 1974; Plumb, 1975; Hou et al., 1990; Newman & Leovy, 1990; Takagi & Matsuda, 2007], in which the zonal wind speed of the atmosphere at the cloud top (~70 km) is more than 60 times faster than the rotation speed of the solid body of Venus. The structures of the thermal tides have been confirmed in a temperature field from space and ground-based observations [e.g., Ainsworth and Herman, 1978; Apt et al., 1980; Taylor et al., 1980; Zasova et al., 2007; Migliorini et al., 2012] and in zonal and meridional wind fields at the cloud top level by tracking cloud motions [e.g., Limaye & Suomi, 1982; Rossow et al., 1990; Sanchez-Levega et al., 2008; Moissle et al., 2009; Kouyama et al., 2012; Horinouchi et al., 2018]. However, despite the importance of thermal tides in the Venusian atmosphere, the global feature of thermal tides across all local times and latitudes has not been obtained due to the limited data coverage of previous observations (e.g., a ground-based observation did not cover the subsolar region [Apt et al., 1980], cloud tracking was limited to only dayside, and only one hemisphere was observable by spacecraft with a polar orbit [Taylor et al., 1980; Migliorini et al., 2012]). Ground-based observations by Ainsworth and Herman［1978］covered the northern and southern hemispheres simultaneously, but the observational dates were quite limited, resulting large data gaps in the high latitudes. Since the tidal amplitude at middle to high latitudes from their results is much greater than those obtained by space-borne observations, verification is needed.

In this study, latitudinal profiles of diurnal, semidiurnal, and higher-frequency components of thermal tides in the temperature field were investigated based on long-term observation data from Longwave Infrared Camera (LIR) onboard Akatsuski [Taguchi et al., 2007]. LIR detects thermal emission from the Venusian cloud layer (60-70km) and has been observing Venus continuously since December 2015. Thanks to the Akatsuki's equatorial orbit, the whole globe of Venus can be observed. These sufficiently long-term and global observations allow us to study the global structure of thermal tides over the whole local time and for a wide latitudinal range in both hemispheres reliably. Because latitudinal phase variations in the thermal tides indicate horizontal momentum transportation, the global profile of the thermal tides is crucial for understanding its impact in the Venusian atmosphere. In addition to the horizontal structure, this study accesses the vertical structure of the thermal tides through a comparison of the tidal structures at different altitudes, using the emission angle dependence of sensing altitudes. To the best of our knowledge, this paper is the first to discuss the wave type of the thermal tides based on the long-term observational results.

**2 Observations and Data**

LIR covers wavelengths of 8–12 μm and captures thermal emissions from the cloud top level of Venus [Taguchi et al., 2007]. An advantage of LIR is that it can observe both the dayside and nightside of Venus. LIR continuously observes Venus at one or two-hour intervals, resulting in 8-15 observations every Earth day except during the superior conjunction, where the spacecraft cannot communicate with the tracking station on Earth for ~1 month. In this study, we used more than 22,000 LIR images obtained from October 2016 to January 2019 (more than three Venusian years). The spatial resolution of the images of Venus changes according to the distance between Venus and Akatsuki. Near the periapsis, the resolution is better than 10 km/pixel, whereas it reaches more than 300 km/pixel at the apoapsis altitude of ~360,000 km where the radius of the Venus disk is 20 pixels. The LIR images used in this study were obtained from the AKATSUKI Science Data Archive [Murakami et al., 2017].

The noise equivalent temperature difference (NETD) is 0.3 K at the target temperature of 230 K, which corresponds to the temperature resolution of LIR. The absolute temperature uncertainty is 3 K [Fukuhara et al., 2011]. The primary cause of the absolute temperature error is variations of instrument's thermal condition affected by the attitude of the spacecraft. The variation due to the thermal condition was almost equally distributed across all local solar time regions, and thus the effect of the absolute error on the brightness temperature map after averaging (shown later) is expected to be much smaller than the original absolute error of 3K. The relative uncertainty from the NETD is also reduced by the averaging. In total, the temperature uncertainty for tidal components is estimated to be 0.1 K in the present analysis. The LIR images have a background bias that strongly depends on the baffle temperature of LIR. This bias was determined from deep space observations, and successfully removed in all LIR image [Fukuhara et al. 2017]. An additional temperature calibration was performed to reduce the unexpected gradual temperature increase seen in deep space images (Supporting Information, S1).

The contribution function representing the sensing altitude of LIR is centered at 65 km altitude with a full width at half maximum of ~10 km for a nadir viewing geometry (Figure 1). The contribution function was calculated by a line-by-line radiative transfer calculation following Sato et al [2014] with the spectral response of LIR [Taguchi et al. 2007]. The calculation adopted the equatorial (<30°) temperature profile from the Venus International Reference Atmosphere (VIRA) [Seiff et al., 1985], the gaseous absorption profiles from Marcq et al. [2005], the cloud particle number density from Haus et al. [2014], which assumes four different-sized particle modes (mode 1, mode 2, mode 2', and mode 3), composed by 75% $H_2SO_4$ and 25% $H_2O$. Then, the line-by-line results were convolved with the filter transmittance of LIR [Taguchi et al. 2007]

LIR observes higher altitudes for larger emission angles [e.g., Taguchi et al., 2012; Supporting Information, S2]. According to the vertical temperature profile of the Venusian atmosphere (negative lapse rate), the observed brightness temperature becomes cooler at higher emission angles (i.e., limb darkening) in the low and middle latitudes. In other words, the vertical structures of the thermal tides can be investigated by comparing brightness temperature maps from different emission angles, although the wide width of the LIR's contribution function does not allow precise altitude assignments. To investigate the vertical structures of the thermal tides, we selected two specified emission angles: 60° (±3.5°), to obtain a wide latitudinal coverage and 45° (±2.5°) for comparison (Figure 1). The effective altitudes of the contribution function are 68.9 km and 67.6 km for the emission angles of 60° and 45°, respectively, which are evaluated from

$$z_e = \frac{\int z f_e(z) dz}{\int f_e(z) dz} \tag{1}$$

where $z$ denotes the altitude and $f_e$ represents the LIR's contribution function for each emission angle, $e$. This indicates a 1.3 km difference in the sensing altitudes between the two emission angles. The uncertainty of the altitude difference can be less than 0.3 km, which was confirmed by the results comparing with a temperature profile obtained from Akatsuki radio occultation observations [Imamura et al., 2019] (Supporting Information, S2). To obtain a continuous dataset, since the spacecraft was not always above the equator and the latitudinal coverage changes from observation to observation, the latitudinal range of ±55° was analyzed for the emission angle of 60°, while ±40° for 45°.

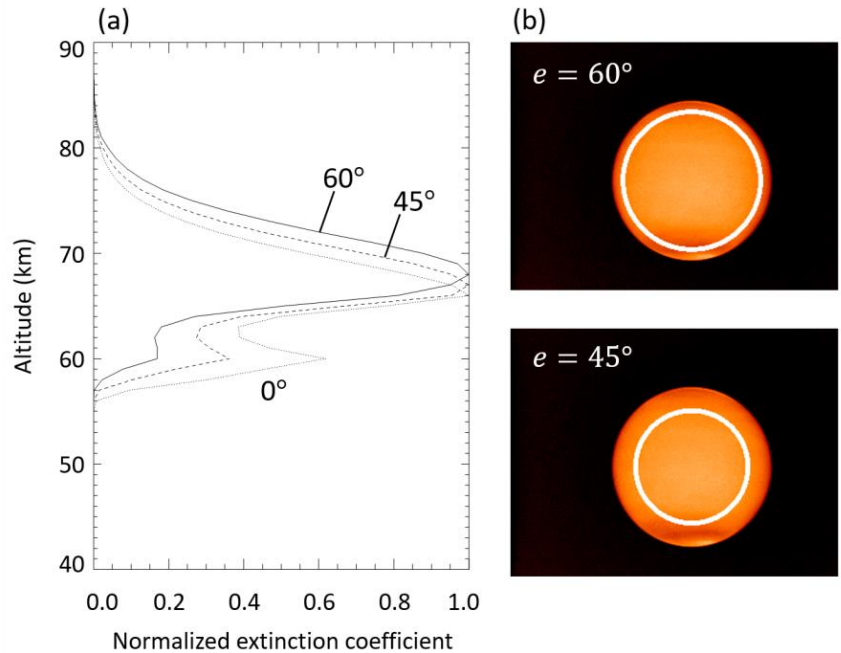

**Figure 1** (a) Contribution functions of the three emission angles, 60° (solid), 45° (dashed), and nadir (dotted), based on the LIR's spectral response. (b) Examples of data coverage in an LIR image for emission angles of 60° and 45°, whose positions are indicated by white circles.

## 3 Global structure of thermal tides

To focus on the solar fixed component in the temperature field, we first averaged the observed brightness temperature as a function of the local time and latitude across the whole observation period. Then, the temperature perturbation, which should mainly consist of thermal tides, was extracted by subtracting the zonal mean brightness temperature at each of the 5-degree latitudinal bins. Figure 2a shows the global structure of the thermal tides without any local time gaps, which was obtained for the first time in the history of Venus observations. Equatorial symmetric structures clearly appeared at all local times. Notably, a local temperature maximum was observed at approximately 9 h in the equatorial region, not at the local noon where solar incident energy is at a maximum. Local peaks at approximately 19 h were observed at the mid-latitudes of both hemispheres. It is worth mentioning that the tidal structure was almost the same for every Venusian year, especially at latitudes lower than 40° (Figure 2b-d). The structures shown in Figure 2a can be considered typical structures of the thermal tides in the Venusian atmosphere during, at least, the Akatsuki observation period.

It should be noted that sensed altitude perturbation can also affect the observed temperature in addition to the atmospheric temperature perturbation, and they cannot be distinguished by the LIR observation alone. In this study we assumed that the temperature fields in Figure 2 were from only atmospheric temperature perturbation, since an effect from the altitude perturbation due to a tidal component can be reduced by the integration effect of the LIR contribution function (the magnitude can be 0.1 K, Supporting Information, S3). This assumption should be validated by a combination of different observations, such as radio occultation and mid-infrared spectral observations in future works.

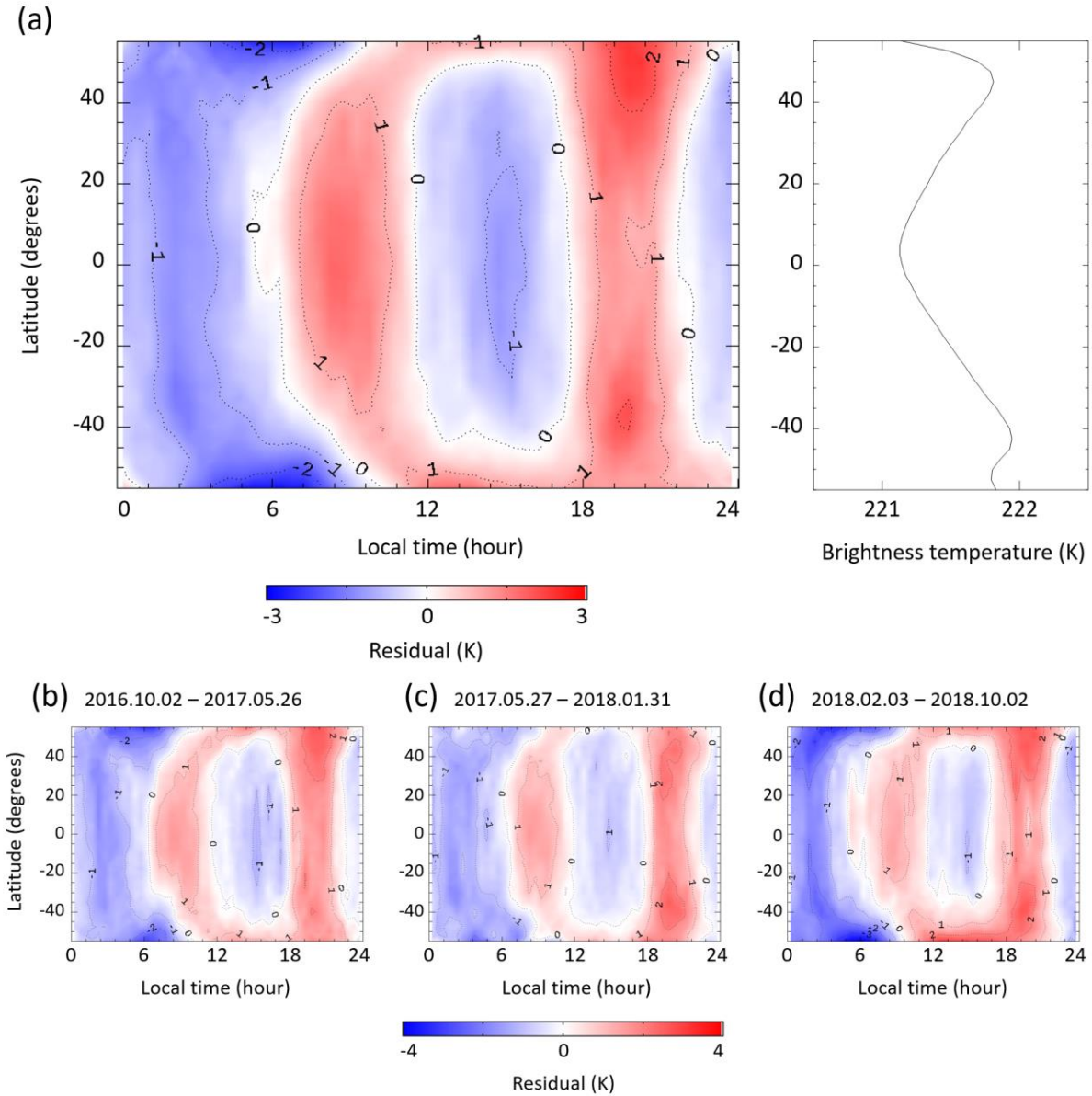

**Figure 2**. (a) Thermal tide structure obtained with a 60° emission angle condition by using the whole data period from October 2016 to January 2019. The brightness temperature in the LIR images was averaged for every 0.5 h local time bin and 5° latitudinal bin, and then the zonal mean temperature was subtracted (right panel of (a)) at each latitude to extract the temperature perturbation. The typical standard error for a grid was 0.08 K. (b-d) Same as (a) but for each Venusian year over three Venusian years.

To clarify which tidal component contributed to the average structure, latitudinal profiles of amplitudes for wavenumber 1-4 components, namely, the diurnal, semidiurnal, terdiurnal, and quarter-diurnal tides, were investigated by fitting sinusoidal functions of wavenumbers 1-4 as

$$f = A_1\sin(x + \varphi_1) + A_2\sin(2x + \varphi_2) + A_3\sin(3x + \varphi_3) + A_4\sin(4x + \varphi_4) \qquad (2)$$

with a nonlinear least squares method (Figure 3), where $x$ denotes local time, and $A_1$-$A_4$ and $\varphi_1$-$\varphi_4$ are fitted parameters. The error in each amplitude was less than 0.05 K evaluated from the fitting. In the low latitudes, the semidiurnal tide was the most significant component with an amplitude greater than 1 K, and its amplitude remained to ~40° latitude. On the other hand, the diurnal tide became significant at middle to higher latitudes where the amplitude of the semidiurnal tide decreased. These tendencies were the same as those at the cloud top level derived from ground-based observations in 1977-1979 [Apt et al., 1980] and the Pioneer Venus Orbiter Infrared Radiometer (OIR) observation in the northern hemisphere [Taylor et al,.1980], indicating long-term steadiness of the thermal tide structure. Note that the amplitude observed by LIR were almost a half of those in the OIR observations. In addition to possible temporal variations in the amplitudes, the difference could be from LIR's long-term observations, which may be more suitable to catch mean structures than the previous observation (ten weeks); however, the LIR's wider contribution function may also affect the amplitude, which integrates the vertical structure of a vertically tilting tidal component and thus smoothens the observed tidal structure. From a simple test for evaluating the integration effect, the actual amplitude of the semi-diurnal tide could be twice as large as the observed amplitude (Supporting Information, S3).

The terdiurnal tide showed small peaks in the mid-latitudes of both hemispheres, which is consistent with a global circulation model (GCM) [Takagi et al., 2018]. Whole components showed almost hemispherical symmetry, but the amplitude of the semidiurnal tide in the northern hemisphere was slightly stronger than that in the southern hemisphere. Recently an asymmetric structure was reported in zonal winds at the cloud top (Horinouchi et al., 2018). There could be a mechanism that induces the asymmetric condition in the Venusian atmosphere whose impact is worth investigating with a numerical approach.

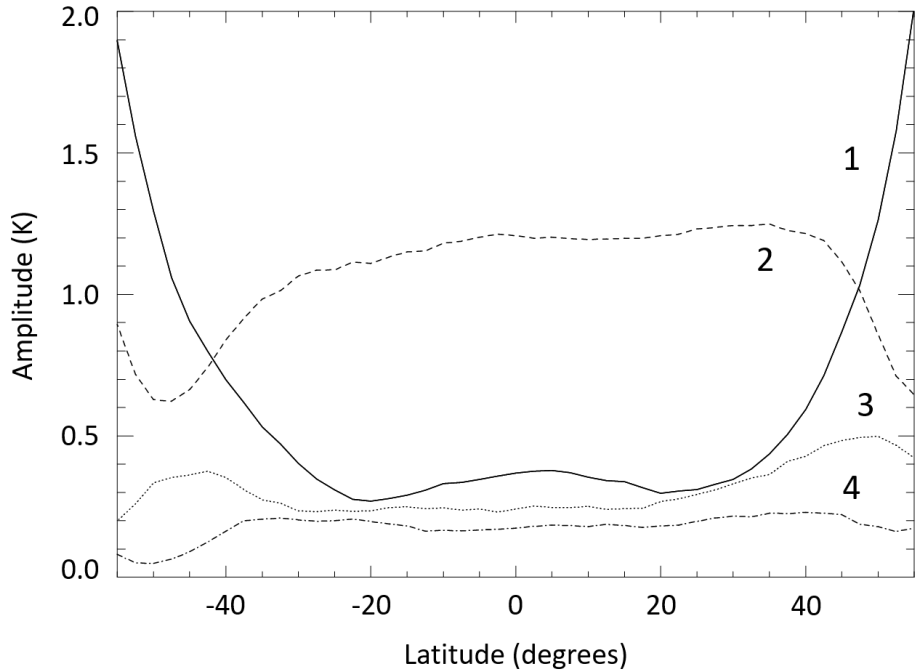

**Figure 3.** Latitudinal amplitude profiles of diurnal (solid line), semidiurnal (dashed), terdiurnal (dotted), and quarter-diurnal (dash-dotted) tides in the 60° emission-angle dataset.

The analysis of the tidal components also provides latitudinal profiles of phases from which a horizontal structure of each tidal component can be reproduced (Figure 4). The diurnal tide showed a clear latitudinal phase tilt toward the evening (westward) direction from the equator to the mid-latitudes. The semidiurnal tide had a flat phase profile from the low to middle latitudes, which was in good agreement with the latitudinal range of the significant amplitude. On the other hand, the semidiurnal tide showed a clear phase tilt toward the evening direction around the mid-latitudes.

The local maximum of the semidiurnal tide clearly shifted from noon, whereas the diurnal tide had a local maximum approximately at noon in the equatorial region. Since solar heating is at a maximum at noon and it is uniformly zero on the night side, the components of both wavenumbers 1 and 2 of the heating should have their local maxima at noon at their excitation altitude. Therefore, the phase shift observed in the semidiurnal tide indicates vertical propagation of the wave from its excitation altitude, although the excitation altitude has not yet been determined.

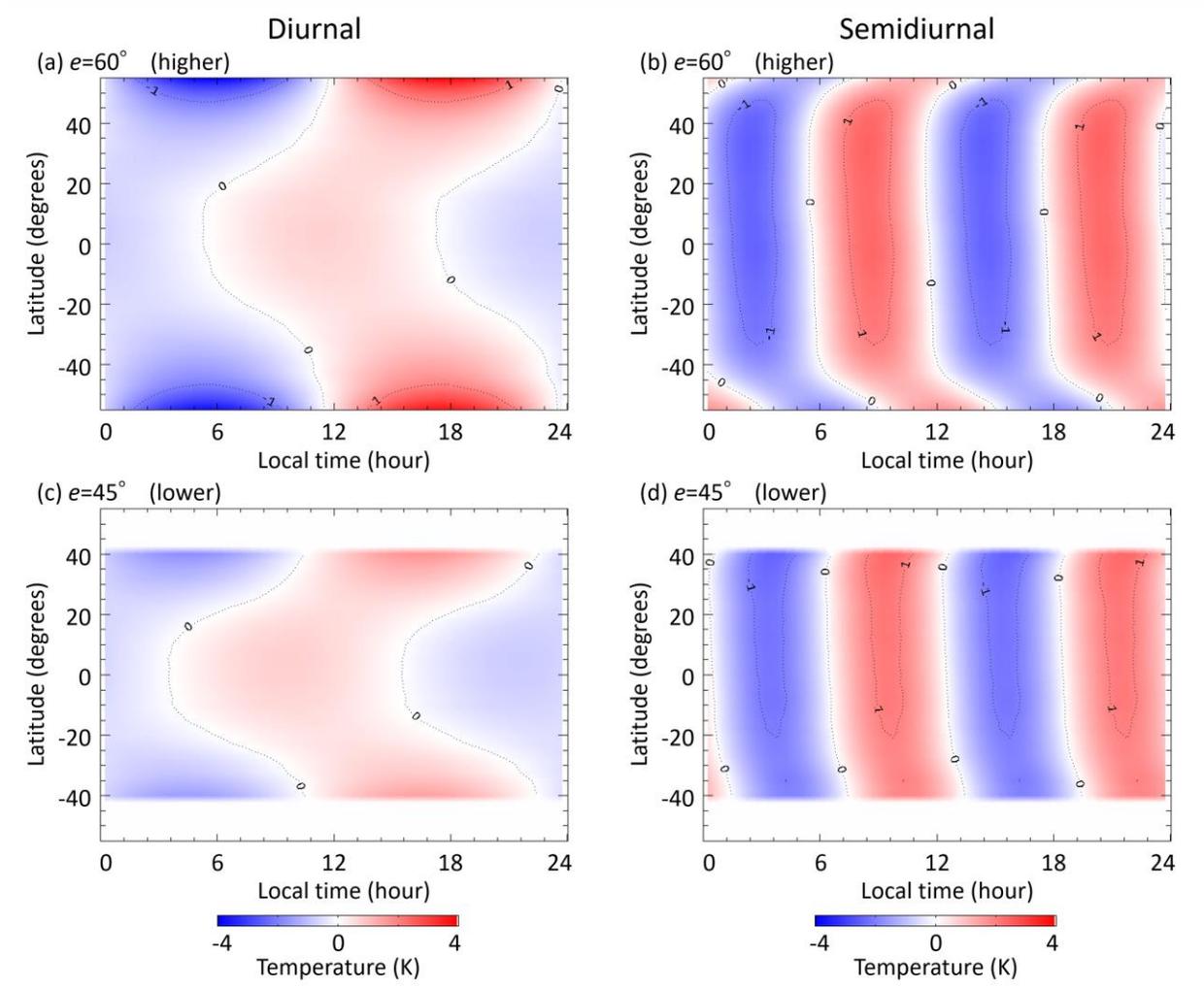

**Figure 4.** Horizontal structures of (a) diurnal and (b) semidiurnal components obtained with the 60° emission angle (higher altitudes) and (c, d) those with the 45° emission angle (lower altitudes). Note that the direction of the superrotation is from morning to evening.

## 4 Discussion

### 4.1 Wave modes of thermal tides and their implications

The classical tidal theory based on Laplace's tidal equation [Longuet-Higgins, 1968] is useful to interpret the observed tidal waves in this study. The superrotating zonal winds act as the background rotation for the thermal tides. If we assume solid-body background rotation, the non-dimensional parameters $\sigma$ and $\gamma$ introduced by Longuet-Higgins [1968] have the same values for each of the diurnal or the semidiurnal tide, irrespective of the planet (whether on Venus or on Earth); see the Supporting Information (S4) for the definition of the parameters and the evaluation of the dimensional parameters. The numerical study by Takagi and Matsuda [2005] suggests that Venusian tidal waves on realistic background winds are somewhat similar to those on the uniform rotation, so the actual tidal waves can be interpreted as modifications of the tidal modes in the theory of Longuet-Higgins [1968].

From the tidal theory, the diurnal tidal features in Figure 4a and 4c are interpreted as the superposition of (mainly) two modes: the gravest symmetric Rossby-wave mode with a negative equivalent depth and the gravest symmetric gravity-wave mode, as is the case for the atmospheric thermal tides on Earth [e.g., Chapman and Lindzen, 1970]. The diurnal Rossby-wave mode has a large amplitude at the middle to high latitudes. The mode is vertically evanescent, so the temperature disturbance should be nearly in phase with the geopotential height disturbance. In fact, the observed cloud top meridional winds of thermal tides [Horinouchi et al., 2018] are consistent with this temperature-geopotential relation. The vertical evanescence is also indicated from the tidal structure in Figure 4a and 4c (see section 4.2). The diurnal gravity-wave mode has an equivalent depth of ~40 m, corresponding to a vertical wavelength of 6 km (Supporting Information, S4). Such a short vertical wavelength was confirmed in a linearized calculation [Pechmann and Ingersoll, 1984]. The corresponding equatorial radius of deformation is $\sim 2.5 \times 10^3$ km, so the mode has nodes of temperature and geopotential height at 25-30°N/S. Then, the amplitude at higher latitudes should be quite small that was inconsistent from the observation, which is more evidence that the diurnal tide cannot be interpreted solely as a gravity wave.

The observed structure of the diurnal tide is possibly superposition of the two wave modes, and such superposition may cause a complex phase relationship between temperature and geopotential height. Therefore, it is difficult to assess momentum transport due to the diurnal tide from the observed latitudinal phase variation of the diurnal component even qualitatively. Moreover, the short vertical wavelength of the gravity wave mode introduces an additional complexity for evaluating momentum transportation due to the diurnal tide discussed in section 4.2. However, one can expect that the diurnal tide may carry the westward (along with the superrotation) momentum toward low latitudes since its thermal forcing mainly resides at low latitudes.

The solution for the semidiurnal tide in the tidal theory consists only of gravity waves. The semidiurnal temperature disturbances in Figure 4b and 4d indicate the gravest symmetric gravity-wave mode. This mode is expected to have a vertical wavelength of 22 km, and the corresponding equatorial deformation is over 40° (Supporting Information, S4), which is consistent with the observed flat structure of the semidiurnal tide. It is rather difficult to interpret the phase tilt at mid-latitudes in Figure 4b, because it may reflect a gradual decrease in the cloud top level from middle to high latitudes [Ignatiev et al., 2009] and a mid-latitude jet structure.

### 4.2 Vertical propagation of the tides

The emission angle dependence of the effective altitude enables us to investigate the vertical structure of the tidal waves. In the diurnal tide (Figure 4a and 4c), there is no apparent phase difference at the mid-latitudes between the emission angles of 45° and 60°. This feature is consistent with the vertical evanescence of the Rossby-wave mode. At low latitudes, there exists a phase difference consistent with the westward tilt (toward evening side) with altitude. This result indicates downward energy (group) propagation. However, since the depth of the solar heating should be much broader than half of the theoretical vertical wavelength, 6 km, a strong superposition is expected to arise from the vertical heating structure. Therefore, it is difficult to conclude whether the wave is truly propagating downward from the cloud-top altitude. Nevertheless, if we assume an altitude difference of 1.3 km between the two sensing altitudes (higher for the emission angle of 60° and lower for the emission angle of 45°) and if we tentatively assume a vertically monochromatic wave near the equator, the phase difference indicates that the vertical wavelength is ~10 km. Note that the vertical superposition owing to the heating distribution is the likely cause of the vertical evanescence of the simulated diurnal tide around the cloud top level even near the equator according to the GCM [Takagi et al., 2018].

On the other hand, the phase difference of the semidiurnal tide had the eastward tilt (toward morning side) with altitude, which is consistent with the direction of upward energy propagation. Quantitatively, the phase difference was 0.86 h, which indicates that the vertical wavelength of the semidiurnal tide was 18 ± 4 km if we assume a 1.3 ± 0.3 km altitude difference between the two sensed altitudes, and this wavelength was almost the same as the estimation from radio occultation observations [Ando et al., 2019]. In addition, the measured vertical wavelength is also consistent with the theoretical expectation of 22 km. Therefore, the actual group propagation is likely upward. Further study is needed to elucidate the full emission-angle dependency of the thermal tide phases.

Vertical propagation of the semidiurnal tide at the cloud level was reproduced in several GCMs [Lebonnois et al., 2016; Takagi et al., 2018; Yamamoto et al., 2019]. However, the phases of the modeled semidiurnal tides at the cloud level were different in different models, and some model results were different from the LIR result. For example, Takagi et al. [2018] showed that the semidiurnal tide has a local temperature maximum of approximately 15 h at the cloud level in their model, whereas our result suggests the local temperature maxima around 9 h and 21 h. Considering the vertical propagation of the semidiurnal tide, the difference may provide a constraint for the solar heating profile in a model, because the wave phase at an altitude should be sensitive to the excitation altitude of the wave that is affected from the solar heating profile.

## 5 Conclusions

In this study, we presented a global structure of thermal tides in the upper cloud layer of Venus using LIR's long-term data, acquired from Akatsuki's equatorial orbit. This global structure was revealed for the first time in the history of Venus ground and space-based observations. The observed structure of the thermal tides showed a good consistency with previous observations, indicating the steadiness of the tides in the Venusian atmosphere. There were fewer variations in the thermal tide structure during the three Venusian years analyzed in this study.

Tidal components were investigated with a periodical analysis. The extracted structure of the diurnal tide, which had a small amplitude at low latitudes and a larger amplitude at higher latitudes, indicated a superposition of the gravest symmetric Rossby-wave and gravity-wave modes. On the other hand, the semidiurnal tide component was significant in the lower to mid-latitudes, and its flat phase structure was consistent with the gravest symmetric gravity-wave mode. The phase difference in the semidiurnal structures between the emission angles of 45° and 60° was consistent with upward energy propagation and a vertical wavelength of ~18 km, which is close to the theoretically expected value. A similar emission-angle comparison for the diurnal tide suggested vertical evanescence in the mid-latitudes, which is also consistent with the tidal theory.

Because recent observations of Venus suggest a global albedo variation for Venus [Lee et al., 2015; Lee e al., 2019] that should affect the excitation of thermal tides, longer-term LIR data will be useful in monitoring possible temporal variation in the thermal tides. Since LIR observations can allow to retrieve the vertical structure of the atmosphere by utilizing the emission angle dependence of sensing altitude, a combination of such LIR observations and radio occultation observations [cf. Ando et al., 2017] will clarify the excitation altitude of the tides and their vertical propagations. By combining the LIR results with numerical modeling, it is also possible to deduce the vertical structure of heating that excites thermal tides, which would greatly advance our understanding of the dynamics of the Venusian atmosphere. In addition, the thermal tide structure derived in this study may provide a more realistic three-dimensional atmospheric condition that will help to discuss a local temperature profile for small scale convection activity [cf. Lefevre et al., 2018].

**Acknowledgments and Data**

This study is supported by the following grants: JSPS KAKENHI 16H02225, 16H02231, 16K17816, 19K14789, EU H2020 MSCA No. 841432, and the MEXT-Supported Program for the Strategic Research Foundation at Private Universities (S1411024). The authors thank Dr. Ando for the valuable information for managing RS data. The version of the LIR and the RS data used in this study is v20190301 and the data will be available at the AKATSUKI Science Data Archive by June 2020 and at the PDS of NASA (https://pds.nasa.gov).

**References**

Ainsworth, J. E., and Herman, J. R. (1978). An Analysis of the Venus Thermal Infrared Temperature Maps. *Journal of Geophysical Research*, **83**, 3113-3124.

Ando, H., Takagi, M., Fukuhara, T., Imamura, T., Sugimoto, N., Sagawa., H., Noguchi, K., Tellmann, S., et al. (2019). Local Time Dependence of the Thermal Structure in the Venusian Equatorial Upper Atmosphere: Comparison of Akatsuki Radio Occultation Measurements and GCM Results. *Journal of Geophysical Research,* **123**, 2270-2280, https://doi.org/10.1029/2018JE005640

Andrews, D. G., Holton, J. R., and Leovy, C. B. (1987). Middle atmosphere dynamics, Academic Press, San Diego, New York, Boston, London, Sydney, Tokyo, Toronto.

Apt, J., Brown, R. A., and Goody R. M. (1980). The character of the thermal emission from Venus. *Journal of Geophysical Research*, **85**, 7934-7940.

Chapman, S., and Lindzen, R. S. (1970). *Atmospheric Tides*, 200 pp. Dordrecht, Reidel.

Fels, S. B. and Lindzen, R. S. (1974). The interaction of thermally excited gravity waves with mean flows, *Geophysical and Astrophysical Fluid Dynamics*, **6**, 149-191.

Fukuhara, T., Taguchi, M., Imamura, T., Nakamura, M., Ueno, M., Suzuki, M., Iwagami, N. et al. (2011). LIR: Longwave Infrared Camera onboard the Venus Orbiter Akatsuki. *Earth, Planets and Space*, 63(9), 1009-1018. https://doi.org/10.5047/eps.2011.06.019

Fukuhara, T., Taguchi, M., Imamura, T., Hayashitani, A., Yamada, T., Futaguchi, M., Kouyama, T., Sato, T. M., et al. (2017). Absolute calibration of brightness temperature of the Venus disk observed by the Longwave Infrared Camera onboard Akatsuki. *Earth, Planets and Space*, 69:141. DOI 10.1186/s40623-017-0727-y

Haus, R., Kappel, D., and Arnold, G. (2014). Atmospheric thermal structure and cloud features in the southern hemisphere of Venus as retrieved from VIRTIS/VEX radiation measurements. *Icarus*, **232**, 232-248.

Horinouchi, T., Kouyama, T., Lee, Y. J., Murakami, S., Ogohara, K., Takagi, M., Imamura, T., Nakajima, K., et al. (2018). Mean winds at the cloud top of Venus obtained from two-wavelength UV imaging by Akatsuki, *Earth, Planets, and Space*, 70:10. https://doi.org/10.1186/s40623-017-0775-3

Hou, A. Y., Fels, S. B., and Goody, R. M. (1990). Zonal superrotation above Venus' cloud base induced by the semidiurnal tide and the mean meridional circulation. *Journal of Atmospheric Sciences*, **47**, 1894–1901

Ignatiev, N. I., Titov, D. V., Piccioni, G., Drossart, P., Markievicz, W. J., Cottini, V., Roatsch, Th., Almedia, M., et al. (2009). Altimetry of the Venus cloud tops from the Venus Express observations. *Journal of Geophysical Research*, **114**, E00B43. doi:10.1029/2008JE003320

Imamura, T., Ando, H., Tellmann, S., Patzold, M., Hausler, B., Yamazaki, A., Sato, T. M., Noguchi, K., et al. (2017). Initial performance of the radio occultation experiment in the Venus orbiter mission Akatsuki. *Earth, Planets and Space*, 69:137. https://doi.org/10.1186/s40623-017-0722-3

Kouyama, T., Imamura, T., Nakamura, M., Satoh, T., and Futaana, Y. (2012). Horizontal structure of planetary-scale waves at the cloud top of Venus deduced from Galileo SSI images with an improved cloud-tracking technique. *Planetary and Space Science*, **60**, 207–216. https://doi.org/10.1016/j.pss.2011.08.008

Lebonnois, S., Sugimoto, N., and Gilli, G. (2016). Wave analysis in the atmosphere of Venus below 100-kmaltitude, simulated by the LMD Venus GCM. *Icarus*, **278**, 38–51. https://doi.org/10.1016/j.icarus.2016.06.004

Lee, Y. J., Imamura, T., Schroder, S. E., and Marq, E. (2015). Long-term variations of the UV contrast on Venus observed by the Venus Monitoring Camera on board Venus Express, *Icarus*, **253**, 1-15.

Lee, Y. J., Jessup, K. L., Perez-Hoyos, S., Titov, D. V., Lebonnois, S., Peralta, J., Horinouchi, T., Imamura, T. et al., Long-term variations of Venus' 365-nm albedo observed by Venus Express, Akatsuki, MESSENGER, and Hubble Space Telescope, *Astronomical Journal*, accepted.

Lefevre, M., Lebonnois, S., and Spiga, A. (2018). Three-Dimensional Turbulence-Resolving Modeling of the Venusian Cloud Layer and Induced Gravity Waves: Inclusion of Complete Radiative Transfer and Wind Shear, *Journal of Geophysical Research*, **123**, 2773–2789. https://doi.org/10.1029/2018JE005679

Limaye, S. S., and Suomi, V. E. (1981). Cloud motions of Venus: Global structure and organization. *Journal of Atmospheric Sciences*, **38**, 1220–1235.

Longuet-Higgins, M. S. (1968). The eigenfunction of Laplace's tidal equations over a sphere. *Mathematical and Physical Sciences*, **262**, 511-607.

Marq, E., Bezard, B., Encrenaz, T., and Birlan, M. (2005). Latitudinal variations of CO and OCS in the lower atmosphere of Venus from near-infrared nightside spectro-imaging. *Icarus*, **179**, 375-386.

Migliorini, A., Grassi, D., Montabone, L., Lebonnois, S., Drossart, P., and Piccioni, G. (2012), Investigation of air temperature on the nightside of Venus derived from VIRTIS-H on board Venus-Express. *Icarus*, **217**, 640-647.

Moissl, R., Khatuntsev, I., Limaye, S. S., Titov, D. V., Markiewicz, W. J., Ignatiev, N. I., Roatsch, T., Matz, K.-D., et al. (2009). Venus cloud top winds from tracking UV features in Venus Monitoring Camera images. *Journal of Geophysical Research*, **114**, E00B31. https://doi.org/10.1029/2008JE003117

Murakami, S., Kouyama, T., Fukuhara, T., Taguchi, M., McGouldrick, K., Yamamoto, Y., and Hashimoto, G. L. (2017). Venus Climate Orbiter Akatsuki LIR Calibrated Data (Version 1.0) [Data set]. Institute of Space and Astronautical Science, Japan Aerospace Exploration Agency. https://doi.org/10.17597/ISAS.DARTS/VCO-00012

Newman, M., and Leovy, C. (1992). Maintenance of strong rotational winds in Venus' middle atmosphere by thermal tides. *Science*, **257**, 647–650.

Pechmann, J. B., and Ingersoll, A. P. (1984). Thermal tides in the atmosphere of Venus: Comparison of model results with observations. *Journal of Atmospheric Sciences*, **41**, 3290–3313.

Plumb, R. A. (1975). Momentum transport by the thermal tide in the stratosphere of Venus. *Quarterly Journal of the Royal Meteorological Society*, **101**, 763–776.

Rossow, W. B., Del Genio, A. D., and Eichler, T. P. (1990). Cloud-tracked winds from Pioneer Venus OCPP Images. *Journal of Atmospheric Sciences*, **47**, 2053–2084.

Sánchez-Lavega, A., Hueso, R., Piccioni, G., Drossart, P., Peralta, J., Pérez-Hoyos, S., Wilson, S. F., Taylor, F. W., et al. (2008). Variable winds on Venus mapped in three dimensions. *Geophysical Research Letters*, **35**, L13204. https://doi.org/10.1029/2008GL033817

Sato, T. M., Sagawa, H., Kouyama, T., Mitsuyama, K., Satoh, T., Ohtsuki, S., Ueno, M., Kasaba, Y., et al. (2014). Cloud top structure of Venus revealed by Subaru/COMICS mid-infrared images. *Icarus*, **243**, 386-399.

Seiff, A., Schofield, J. T., Kilore, A. J., Taylor, F. W., Limaye, S. S., Revercomb, H. E., Sromovsky, L. A., Kerzhanovich, V. V., et al. (1985). Models of the structure of the atmosphere of Venus from the surface to 100 kilometers altitude. *Advanced in Space Research*, **5**, 3-58.

Taguchi, M., Fukuhara, T., Imamura, T., Nakamura, M., Iwagami, N., Ueno, M., Suzuki, M., Hashimoto, G. L., et al. (2007). Longwave Infrared Camera onboard the Venus Climate Orbiter. *Advances in Space Research*, **40**, 861-868.

Taguchi, M., Fukuhara, T., Futaguchi, M., Sato, M., Imamura, T., Mitsuyama, K., Nakamura, M., Ueno, M., et al. (2012). Characteristic features in Venus' nightside cloud-top temperature obtained by Akatsuki/LIR. *Icarus*, **219**, 502-504.

Takagi, M. & Matsuda Y. (2005). Sensitivity of thermal tides in the Venus atmosphere to basic zonal flow and Newtonian cooling, *Geophysical Research Letters*, **32**, L02203. https://doi.org/10.1029/2004GL022060

Takagi, M., and Matsuda Y. (2007). Effects of thermal tides on the Venus atmospheric superrotation. *Journal of Geophysical Research: Atmospheres*, **112**, D09112. https://doi.org/10.1029/2006JD007901

Takagi, M., Sugimoto, N., Ando, H., and Matsuda, Y. (2018). Three-Dimensional Structures of Thermal Tides Simulated by a Venus GCM. *Journal of Geophysical Research: Planets*, **123**, 335-352. https://doi.org/10.1002/2017JE005449

Yamamoto, M., Ikeda, K., Takahashi, M., and Horinouchi, T. (2019). Solar-locked and geographical atmospheric structures inferred from a Venus general circulation model with radiative transfer. Icarus, 321, 232-250. https://doi.org/10.1016/j.icarus.2018.11.015

Zasova, L. V., Ignatiev, N. I., Khatountsev, I. V., & Linkin, V. (2007). Structure of the Venus atmosphere. *Planetary and Space Science*, **55**, 1712–1728.

## Supporting Information

**Introduction**

This Supporting Information provides the calibration procedure for the reducing unexpected temperature increase observed in the LIR data (S1) even after applying the absolute temperature correction described by Fukuhara et al. [2017]. This Supporting Information also provides the cause of the altitude difference at different emission angles and the uncertainty of the estimated difference with the model we used in this study (S2). Then, this Supporting Information shows an experimental evaluation of the integration effect of the LIR contribution function. Finally, this Supporting Information provides theoretical explanations of the tidal structures in the Venusian atmosphere based on those from Longuet-Higgins [1968] (S4).

**S1 Correction for the unexpected increase in the brightness temperature**

Since September 2016, LIR has conducted four deep space observations for monitoring the LIR's status, and unexpected temperature increases have been confirmed in the deep space images. The magnitude of the increase reached 10 K in deep space images whose mean temperature in an LIR image frame was 180-190 K. Here, 180 K is a lower detection limit of the LIR measurement. The increase of 10 K at 180 K corresponds to 3 K at 230 K, which is a typical temperature at the cloud top level. Because deep space is an ideal calibration target that has no energy input to the LIR, the temperature increase should be the LIR's instrumental calibration issue.

Similar to a procedure used by Fukuhara et al. [2017], we first converted the observed brightness temperature to the corresponding brightness by assuming black-body radiation. Before the conversion, a correction in the absolute temperature described by Fukuhara et al. [2017] was performed. As shown in Figure S1, there was a linear increasing trend in the mean brightness. The non-zero brightness for deep space comes from the lower detection limit of the LIR. On the other hand, there was no clear temporal variation in the standard deviation of the brightness, and the standard deviation was much smaller than the magnitude of the increase. This finding indicates that the increase occurred for all pixels, and the magnitude of the increase was the same at any position in the LIR image frame. These tendencies indicate that the increase can be considered an offset variation. We evaluated the magnitude of the increase in the brightness with a linear model

(Figure S1a), and then, we subtracted the brightness increase from each LIR image; the temperature increase seen in deep space images was successfully reduced (Figure S1b).

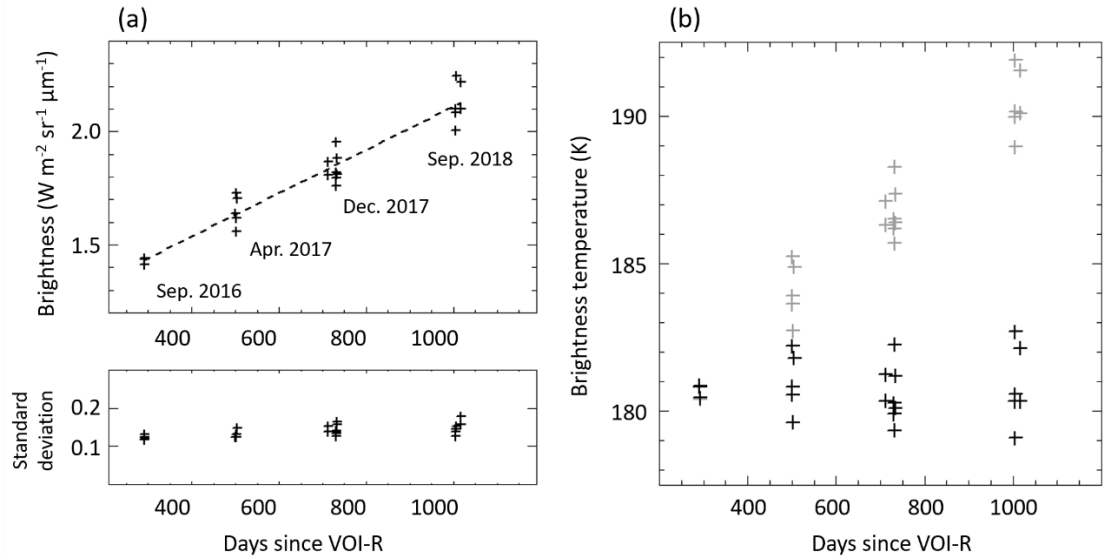

**Figure S1.** (a) Mean brightness of the LIR deep space images and their standard deviations. In each observation campaign, the LIR took several deep space images with different baffle temperatures. In this plot, we used only images with a baffle temperature cooler than 30 °C. (b) Brightness temperature from the deep space images before correcting the temperature increase (gray) and after the correction (black).

## S2. Emission angle dependence of the LIR's sensing altitude and its uncertainty

As shown in Figure S2a, which shows an ideal case where the number density of cloud particles is the same at any altitude, an altitude of $\tau = 1$ seen from LIR increases proportionally to $1 - \cos e$, where $e$ represents an emission angle. This characteristic is basically the same when we consider with a more realistic cloud model; that is, the effective altitude estimated from (1) with the LIR's contribution function increases monotonically (and almost linearly) as a function of $1 - \cos e$ (Figure S2b).

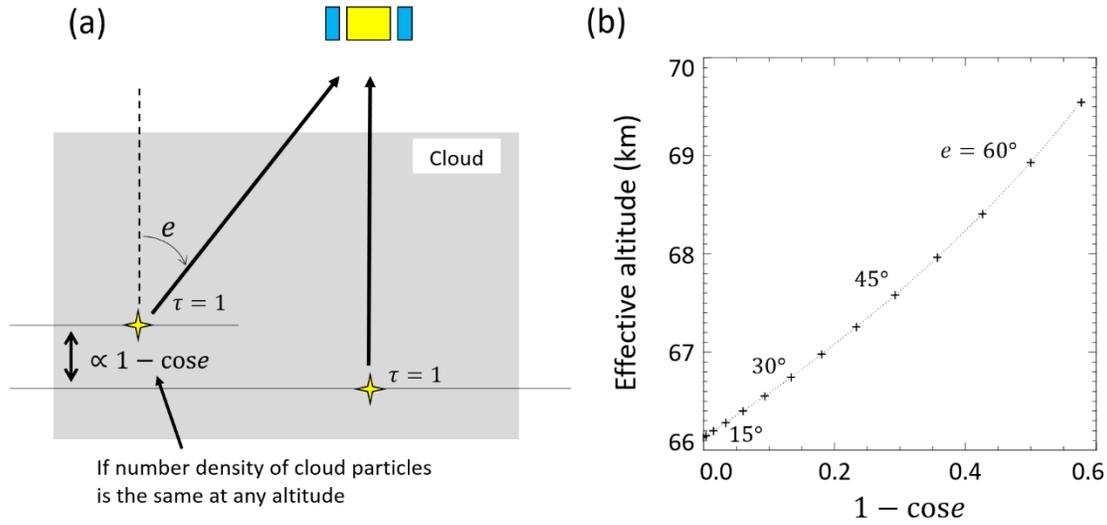

**Figure S2.** (a) Schematic view of the cause of emission angle dependence for the LIR sensing altitude. If the number density of cloud particles is the same at any altitude, an altitude of $\tau = 1$ increases proportionally to $1 - \cos e$. (b) Effective altitudes for various emission angles (5-degree intervals).

Next, we evaluated the reliability of the altitude difference between different emission angles in our analysis. The reliability of the altitude difference is more important in this study for investigating a vertical structure of a tidal component than the absolute altitude. To do this, we estimated the zonal mean brightness temperature at several emission angles, not only 45° and 60°, but we also used 0°, 15°, and 30° (more precisely, we used $e$=60°±3.5°, 45°±2.5°, 30°±2°, 15°±1.5°, and 1.5°±1.5°. The emission angle ranges were adjusted to avoid any data-less latitudes in small Venus disk images), and evaluated the temperature decreasing rate with the five effective altitudes estimated from (1). Then, we compared the obtained rate with those from radio occultation observations, termed RS (Radio Science) (Imamura et al., 2018: Ando et al., 2019), and VIRA (Figure S3).

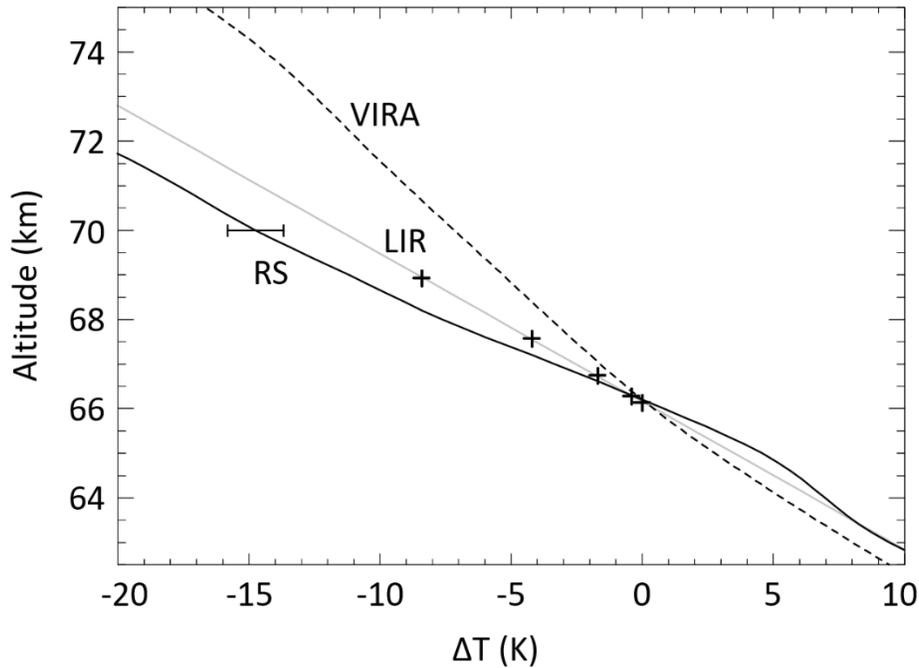

**Figure S3**. Vertical profiles of temperature variations measured from LIR (plus sings) and its fitted line (gray line), Akatsuki/RS (solid line), and VIRA (dotted line). Error at 70 km altitude for the RS data was also shown from a standard error of the RS observations. We used the temperature measured at ~66 km (an effective altitude of e=0° case) as the reference temperature for each profile.

From Figure S3, the LIR temperature variation (-3.0 K km$^{-1}$) is closer to that from the RS profile (-3.4 K km$^{-1}$ on average from 62.5 to 72.5 km) than that from the VIRA profile (-2.0 K km$^{-1}$). Since RS observations were conducted during the LIR observation period, it is natural that the LIR provided a similar temperature variation to that of RS, although there is still a significant gap between the LIR and RS, though there is still a significant gap between LIR and RS. Note that the RS data were obtained at various local times. The difference should reflect the model uncertainty and the temporal variation in the vertical structure of the Venusian atmosphere.

We evaluated the effect from an emission angle range we used to obtain the temperature field in Figure 2. We examined the averaging analysis with several emission angle ranges, that is, 1°, 2.5°, and 3.5° for *e*=60° (in this test, we allowed the situation that sometimes we could not obtain any pixel for the averaging at some latitudes in small Venus disk images). As a result, the zonal mean temperature at the equator was almost same among different emission angle ranges (221.4, 221.2, and 221.0 K for the emission angle ranges of 1°, 2.5°, and 3.5°, respectively). Therefore, we concluded that the effective altitude does not so sensitive to the emission angle range, and we consider that there should be up to 0.4 K uncertainty (= 0.1 km uncertainty in altitude) from the test. Note that the effect from using a wide emission angle range does not produce diurnal and semidiurnal temperature variations like the observed ones, because it only biases temperatures equally for all local times.

Based on the result in Figure S2 and the test for the effect from an emission angle range, we concluded that there could be up to 20% uncertainty in our estimation of altitude difference.

## S3. Evaluating the integration effect of the LIR contribution function

We conducted a simple test to evaluate the integration effect of the LIR contribution function which may smoothen the temperature profile. For the test, we used a temperature profile from VIRA as a base profile and added a semi-diurnal component that propagates from lower to higher altitude (Figure S4). We set the vertical wavelength to 18 km based on the observational result. To simplify, we assumed that the amplitude of the semi-diurnal tide was the same among altitudes.

In the test, we also considered a possible cloud altitude variation induced by the vertical wind perturbation of the semi-diurnal tide. Based on observed tidal signatures in pervious cloud tracking results at the cloud top level (Moissl et al., 2009; Horinouchi et al., 2018), the zonal wind perturbation can be $\hat{u}$ ~ 10 m s$^{-1}$; then, the amplitude of the vertical wind perturbation is $\hat{w}$ ~ 1 cm s$^{-1}$ from the wave's aspect ratio ($\lambda_x$ ~ 20,000 km, while $\lambda_z$ ~ 20 km). Since an air parcel passes one period of the semi-diurnal tide within 2 days because of the fast background zonal flow (circulating with a 4-day period), the amplitude of the cloud altitude variation would be

$$\hat{h} = \frac{1}{2}\int_0^{T/2} \hat{w}\sin\left(\frac{2\pi}{T}t\right) dt = \hat{w}\frac{T}{2\pi} \sim 260 \text{ m}, \tag{S1}$$

where $T$ = 2 days. In the test, the effect was introduced simply by altering the LIR's contribution function up or down according to the phase of the material surface variation.

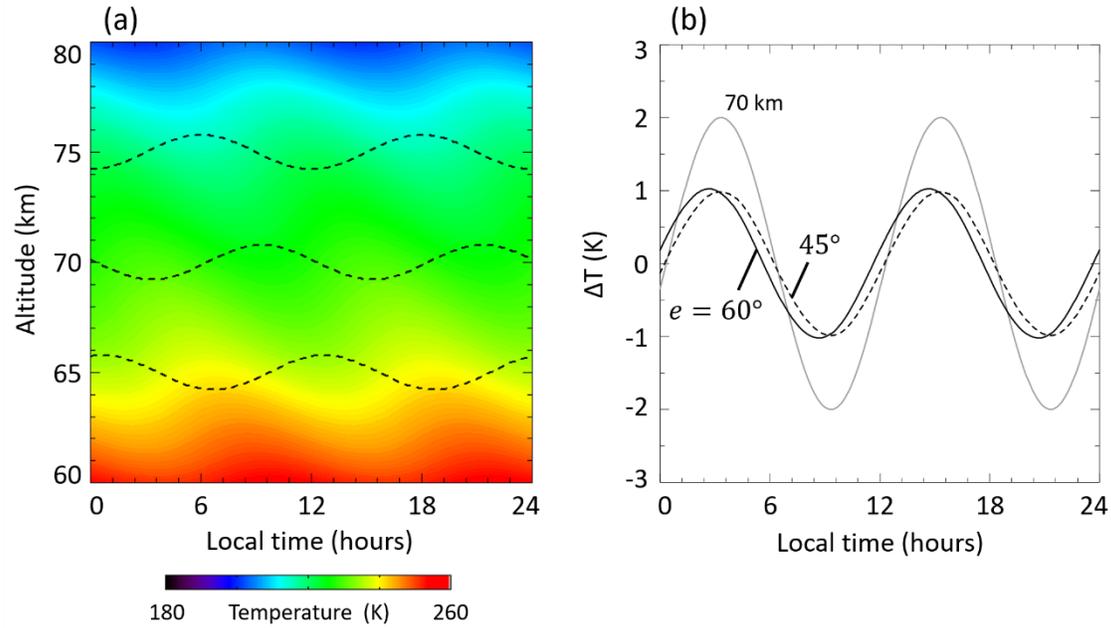

**Figure S4.** (a) Model temperature used in the integration test. Dashed lines indicate phases of cloud altitude variations at 65, 70, and 75 km. (b) Integrated semi-diurnal components with the LIR contribution functions for $e = 60°$ (black solid) and 45° (dashed). Temperature variation at a 70 km altitude is also shown (gray solid).

From the test, the amplitude of the semi-diurnal tide after integrating a vertical temperature profile with the LIR contribution function was almost half that of the given amplitude. Note that we assumed 2 K amplitude in the test by considering the thermodynamic energy equation in a log-p coordinate (cf. Andrews et al., 1987) rotating with the background super-rotation (i.e. $\bar{u} = 0$) with an approximation of adiabatic condition;

$$\frac{\partial T'}{\partial t} = -\frac{N^2 H}{R} w' = -\left(\frac{\partial T}{\partial z} + \frac{\kappa T}{H}\right) w' \tag{S2}$$

where $H$ is a scale height at the cloud level (~5 km), $R$ is the gas constant, $C_p$ is the specific heat at constant pressure, and $\kappa = R/C_p$ (~0.25). Since the period of the semi-diurnal tide can be 2 days in the above equation, we have

$$|T'| \sim \left(\frac{\partial T}{\partial z} + \frac{\kappa T}{H}\right) \frac{t_s}{2\pi} |w'| \sim 2 \text{ K}, \tag{S3}$$

here we used $T$=230 K, $\partial T/\partial z$=-3 K km$^{-1}$, $t_s$=2 days, and $|w'|$=0.01 cm s$^{-1}$. On the other hand, the phase positions for both the $e = 45°$ and $60°$ cases were almost the same as that at a 70 km altitude.

In addition, the phase difference between the two emission angles was clearly observed, although the difference in the test (0.61 h) was somewhat smaller than the observed difference (0.86 h). The difference should reflect uncertainty in our model and the simplified assumption used in the test. However, the direction of the obtained phase difference is consistent with the propagation direction of the given tide.

We also examined another test without the altitude variation due to the tide to understand the effect of the sensed altitude variation. We found the resulting zonal mean temperature showed only a 0.1 K difference from the nominal case.

## S4. Structures of tides based on tidal theory

In this section, we derive theoretical parameter values of thermal tides in the upper cloud layer of Venus. To our knowledge, no such estimation is available in research papers on the tides on Venus.

Super-rotation acts as the background rotation for thermal tides. By regarding it as a solid-body rotation to the first approximation, the parameter values are provided from the numerical solutions of Laplace's tidal equations from Longuet-Higgins [1968, hereinafter LH]. Here, we use the following constants; the planet radius $a = 6,120$ km (considering that the altitude of the upper cloud layer is ~70 km), the gravitational acceleration $g = 9$ m s$^{-2}$, the Brunt-Väisälä frequency $N = 0.02$ s$^{-1}$, and the scale height $H = 5$ km. The angular velocity of the background $\Omega = 2\pi/(4\times86400)$ s$^{-1}$ is based on a superrotation period of 4 days.

The diurnal tide has a zonal wavenumber of 1, and its angular frequency $\sigma$ is equal to $\Omega$. Figure 2 in LH relates the nondimensional angular frequency $\sigma/(2\Omega)$ with another nondimensional parameter called the Lamb parameter, $\gamma$, when it is positive. Here,

$$\gamma \equiv \frac{4\Omega^2 a^2}{gh}, \tag{S4}$$

where $h$ is the equivalent depth. For the gravest-symmetric ($n = 1$) retrograde gravity wave mode, $\gamma \sim (0.09)^{-2}$, then $h \sim 40$ m from the present constants. The equivalent depth is related to the vertical wavelength as (e.g., Andrew et al., [1987])

$$\lambda_z = 2\pi / \sqrt{\frac{N^2}{gh} - \frac{1}{4H^2}}. \tag{S5}$$

When $h = 40$ m, $\lambda_z = 6.3$ km.

Since the equivalent depth of the mode is small, we can further use the equatorial beta approximation to estimate its horizontal structure. This approximation introduces the equatorial deformation radius,

$$l_e \equiv \sqrt{\frac{(gh)^{1/2}}{\beta}}. \tag{S6}$$

From (S6), $l_e$ for the gravity wave mode of the diurnal tide is ~2500 km ($< 25°$ latitude). This result indicates that the temperature disturbance associated with it has a node at around the latitude, and its amplitude attenuates sharply with latitude on its poleward side. For reference, the nondimensional parameter values are independent of planets. The same mode dominates the diurnal thermal tide in the low-latitude middle atmosphere of the Earth, where its vertical wavelength is ~ 30 km (e.g., Andrew et al., [1987]).

Since the diurnal gravity-wave modes are confined to low latitude, the diurnal tide at mid to high latitudes of Venus must be dominated by the gravest symmetric ($n = 1$) Rossby-wave mode with $\gamma \sim 0$ as on Earth (Figure 2 and 17 in LH), which is a vertically evanescent wave.

A non-Rossby-wave solution exists for the semi-diurnal tide (zonal wavenumber 2 and $\sigma = 2\Omega$), so it must consist of gravity waves. The gravest symmetric ($n = 1$) semi-diurnal gravity-wave mode has $\gamma \sim (0.3)^{-2}$ from Figure 3 in LH. Then, from the equations (S4) and (S5), $h =$ ~500 m, and the derived vertical wavelength is ~22 km. The corresponding equatorial deformation radius is ~4700 km ($> 40°$ latitude), which indicates that the semi-diurnal tide has a longer latitudinal extent than that of the diurnal tide. For reference, the same mode dominates the semi-diurnal tide in Earth's middle atmosphere, where its vertical wavelength is on the order of 100 km, so it is virtually evanescent (e.g., Andrew et al., [1987]).